\documentclass[aps,prl,twocolumn,floatfix]{revtex4}
\usepackage{graphicx}
\usepackage{epstopdf}
\usepackage{amssymb}
\usepackage{epsf}
\usepackage{bm}

\newcommand{\bra}[1]{\langle #1|}
\newcommand{\ket}[1]{|#1\rangle}

\renewcommand{\phi}{\varphi}
\renewcommand{\epsilon}{\varepsilon}
\renewcommand{\vec}[1]{{\bf #1}}

\usepackage{amssymb}
\usepackage{amsmath}

\begin{document}
\title{Exploring Topological Phases With Quantum Walks}
\author{Takuya Kitagawa}
\affiliation{Physics Department, Harvard University, Cambridge,
MA 02138, USA}
\author{Mark Rudner}
\affiliation{Physics Department, Harvard University, Cambridge,
MA 02138, USA}
\author{Erez Berg}
\affiliation{Physics Department, Harvard University, Cambridge,
MA 02138, USA}
\author{Eugene Demler}
\affiliation{Physics Department, Harvard University, Cambridge,
MA 02138, USA}

\begin{abstract}
The quantum walk was originally proposed as a
quantum mechanical analogue of the classical random walk, and has since become a powerful tool in quantum information science.
In this paper, we
show that discrete time quantum walks provide a versatile platform for studying topological phases, which are currently the subject of intense theoretical and experimental investigation.
In particular, we demonstrate that recent experimental realizations of quantum walks simulate a non-trivial one dimensional
topological phase.
With simple modifications, the quantum walk can be engineered to realize
all of the topological phases
which have been classified in one and two dimensions.
We further
discuss the existence of robust edge modes at phase boundaries, which provide experimental signatures for the
non-trivial topological character 
of the system.
 \end{abstract}

\date{\today}
\maketitle
Quantum walks, the quantum analogues of classical random walks\cite{originalquantumwalk},
form the basis of efficient quantum algorithms\cite{farhi,kempe}, and provide a universal platform for quantum computation\cite{childs}.
Much like their classical counterparts, quantum walks can be used to model a wide variety of physical processes including photosynthesis\cite{photosynthesis1,photosynthesis2},  quantum diffusion\cite{quantumdiffusion}, optical/spin pumping and vortex transport\cite{rudner}, and electrical breakdown\cite{breakdown1, breakdown2}.
Motivated by the prospect of such an array of applications, several groups have recently realized quantum walks in experiments using ultracold atoms in optical lattices\cite{coldatomexperiment}, trapped ions\cite{ions}, photons\cite{photons}, and nuclear magnetic resonance\cite{NMR}.
These systems offer the possibility to study quantum dynamics of single or many particles in a precisely controlled experimental setting.

Here we show that quantum walks can be used to explore dynamics in
a wide range of topological phases\cite{ssh,shinseizeroenergy,
TKNN}.
Interest in topological phases was first sparked by the
discovery of the integer quantized Hall
effect\cite{integerquantumhall, TKNN},
and has rapidly increased in recent years following the 
prediction\cite{topologicalinsulatorprediction2, topologicalinsulatorprediction1, topologicalinsulatorprediction3} and experimental realization\cite{topologicalinsulatorexperiment1, topologicalinsulatorexperiment2} of a new class of materials 
called ``topological insulators.'' Unlike more familiar
states of matter such as the ferromagnetic and superconducting
phases, which break SU(2) (spin-rotation) and U(1) (gauge)
symmetries, respectively, topological phases do not break any
symmetries and cannot be described by any local order parameters.
Rather, these phases are described by topological invariants which
characterize the global structures of their ground state
wavefunctions. Topological phases are known to host a variety
of exotic phenomena such as fractional charges and magnetic
monopoles \cite{Qi1, Qi2}.

 \begin{figure}
\includegraphics[width = 8.5cm]{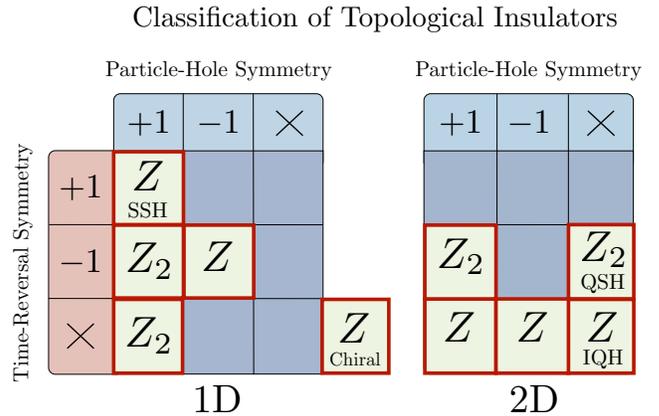}
\caption{Classification of topological phases by symmetry for one (1D) and two (2D) dimensions, adapted
from Ref.\cite{shinseiclassification1, shinseiclassification2}. Discrete time quantum walks
can naturally realize all ten classes of nontrivial
topological phases in 1D and 2D.  
Time reversal symmetry (TRS) and particle-hole symmetry (PHS) are
defined by the existence of antiunitary operators $\mathcal{T}$
and $\mathcal{P}$ satisfying Eqs.(\ref{TRS}) and (\ref{PHS}), 
and
may be absent, or present with $\mathcal{T}^2 = \pm 1$
($\mathcal{P}^2 = \pm 1$). In the absence of both TRS and PHS, a
 distinct ``chiral'' symmetry with a unitary $\Gamma$
satisfying Eq.(\ref{chiral}) 
may be found.
In each case, the symmetry-allowed phases are classified by an
integer ($Z$) or binary ($Z_2$) topological invariant. Classes
containing the Su-Schrieffer-Heeger model (SSH)\cite{ssh}, integer
quantized Hall (IQH)\cite{integerquantumhall,TKNN}, and quantum
spin-Hall (QSH)\cite{topologicalinsulatorprediction2, topologicalinsulatorprediction1, topologicalinsulatorprediction3,
topologicalinsulatorexperiment1, topologicalinsulatorexperiment2} phases are indicated. }
\label{TopTable}
\end{figure}

The class of 
topological phases
which can be realized in a
system of non-interacting particles is determined by the
dimensionality of the system and the underlying symmetries of its
Hamiltonian.
Figure \ref{TopTable} shows the ten classes of topological phases
which can arise in one dimensional (1D) and two dimensional (2D)
systems with and without time-reversal symmetry (TRS)
and particle-hole symmetry (PHS) (see Refs.\cite{shinseiclassification1, shinseiclassification2} and discussion below).
If both symmetries are absent in 1D, the possibility of a distinct ``chiral'' symmetry creates an additional
class of topological phases.
Within each class, the allowed phases are characterized by either an integer ($Z$) or binary ($Z_2$) topological invariant.

In this paper we investigate the topological phases of Fig.\ref{TopTable} in discrete-time quantum walks (DTQWs).
In a DTQW, a walker with a two-fold internal ``spin'' degree of freedom is made to hop between adjacent sites of a lattice through
a series of unitary operations.
We discuss how the DTQW protocol can be engineered to selectively satisfy time-reversal and particle-hole symmetries,
and show that DTQWs can realize 
all of the classes of topological phases in 1D and 2D.
In particular, we show that the DTQWs demonstrated in recent experiments\cite{coldatomexperiment, ions} have already realized a non-trivial one dimensional topological phase, which is analogous to that of 
the Su-Schrieffer-Heeger (SSH) model of polyacetylene\cite{ssh}
(see Fig.\ref{TopTable}).

The non-trivial topological properties of the systems classified in Fig.\ref{TopTable} are manifested in the presence of robust edge states at phase boundaries, i.e. zero energy bound states\cite{shinseizeroenergy} and gapless edge modes\cite{edgestate} in 1D and 2D systems, respectively.
We propose a scheme to identify the presence of topological phases through the observation of edge modes at an interface between regions where different DTQW protocols are applied.

 \begin{figure}
\includegraphics[width = 8.5cm]{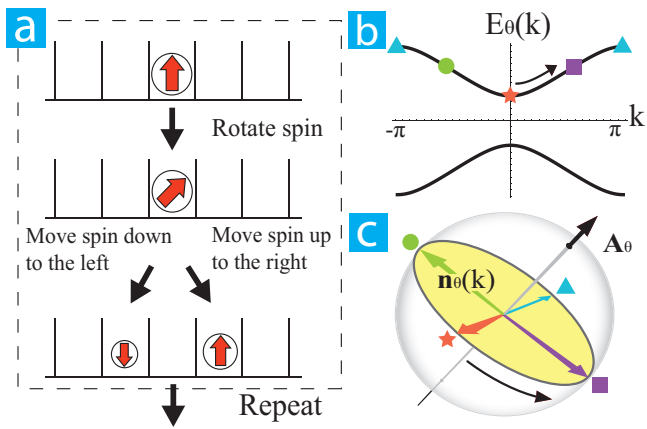}
\caption{(a) One dimensional discrete time quantum walk (DTQW) protocol.
First, the walker's internal ``spin'' is rotated through an angle $\theta$ about the $y$-axis.
Then, the walker is coherently translated by one lattice site to the right (left) if its spin is up (down). 
The quantum walk is produced by repeatedly applying this combined ``step'' operation. 
(b) Band structure of 1D DTQW with $\theta=\pi/2$.
The spinor eigenstates at each momentum $k$ are directed along the unit vector $\vec{n}_\theta(k)$ [see Eq.(\ref{nk})], as represented on the Bloch sphere in panel (c).
Corresponding points in panels (b) and (c) are indicated by the colored markers.
For any $\theta \neq 0, 2\pi$, $\vec{n}_\theta(k)$ winds around the origin once as $k$ traverses the Brillouin zone (black arrow).
The vector $\vec{A}_\theta$ is normal to the plane containing $\vec{n}_\theta(k)$, and defines the axis for the chiral symmetry $\Gamma_\theta$, see Eq.(\ref{chiral}).
}
\label{qwpicture}
\end{figure}

\section{topological phases in 1D}
The one dimensional DTQW protocol employed in recent experiments\cite{coldatomexperiment,ions,photons, NMR} is depicted schematically in Fig.\ref{qwpicture}a. 
The basis states of the system are described in terms of the position of the ``walker,''
defined on integer lattice sites $x$, and its internal ``spin'' state
which can be either up ($\uparrow$) or down ($\downarrow$). 
The quantum evolution is produced by repeatedly applying a unitary operation
\begin{equation}
 \label{BasicWalk}U(\theta) = TR(\theta)
\end{equation}
that defines one ``step'' of the quantum walk. Each step consists of a spin rotation $R(\theta)$, followed
by a coherent spin-dependent translation
\begin{equation}
T = \sum_x \left[ \ket{x+1}\bra{x}\otimes\ket{\uparrow}\bra{\uparrow} + \ket{x-1}\bra{x}\otimes\ket{\downarrow}\bra{\downarrow} \right]
\label{translation}
\end{equation}
that shifts  the walker to the right (left)  by one lattice site if its spin is up (down).
This step protocol is a unitary generalization of the classical process in which a
random walker hops left or right according to the outcome of a
stochastic ``coin-flip.''
Here, as in the experiments of Refs.\cite{coldatomexperiment,ions}, we consider the case where $R(\theta)$ corresponds to
a spin rotation around the $y$-axis through an angle $\theta$,
\begin{equation}
R(\theta)   = \left( \begin{array}{cc} \cos\, (\theta/2)  & -\sin\, (\theta/2) \\ \sin\, (\theta/2) &\ \ \ \cos\, (\theta/2) \end{array} \right).
\label{rotation}
\end{equation}

Although the step protocol is defined explicitly in terms of the discrete unitary operations $T$ and $R(\theta)$,
the net evolution over one step is equivalent to that generated by a time-independent effective Hamiltonian $H(\theta)$ over the step-time $\delta t$,
\begin{equation}
\label{Heff}U(\theta) = e^{-iH(\theta) \delta t},\quad \hbar = 1.
\end{equation}
The evolution operator for $N$ steps is given by $U^N(\theta) = e^{-i H(\theta) N\delta t}$.
Thus, the DTQW provides a stroboscopic simulation of the evolution generated by $H(\theta)$ at the discrete times $N \delta t$.
Below we take units in which $\delta t = 1$.

The DTQW protocol described above is translationally invariant.
The evolution operator $U(\theta) $ and the Hamiltonian $H(\theta) $ are thus diagonalized down to $2\times 2$ blocks in the basis of Fourier modes, $\ket{k}\otimes\ket{\sigma} = \frac{1}{\sqrt{2\pi}} \sum_x e^{-ikx}\ket{x}\otimes\ket{\sigma}$, with $-\pi \le k < \pi$. 
For the choice of $R(\theta)$ in Eq.(\ref{rotation}), $H(\theta)$ can be written as
\begin{equation}
\label{Hk}H(\theta) = \int_{-\pi}^{\pi} dk \left[
E_{\theta}(k) \,\vec{n}_{\theta}(k) \cdot {\bm \sigma}\right]\otimes\ket{k}\bra{k},
\end{equation}
where ${\bm \sigma} = (\sigma_x, \sigma_y, \sigma_z)$ is the vector of Pauli matrices and the unit vector
$\vec{n}_{\theta}(k) = (n_x, n_y, n_z)$ defines the quantization axis for the spinor eigenstates at each momentum $k$.
Because the evolution is prescribed stroboscopically at unit intervals,
the eigenvalues $\pm E_{\theta}(k)$ of $H(\theta)$ are only determined up to integer multiples of $2\pi$.
The corresponding band structure is thus a ``quasi-energy'' spectrum, with $2\pi$ periodicity in energy.
For $\theta \neq 0$ or $2\pi$, 
explicit expressions for $E_{\theta}(k)$ and $n_{\theta}(k)$ are given by
$\cos E_{\theta}(k) = \cos(\theta/2)\,\cos k$ and
\begin{equation}
\label{nk}\vec{n}_{\theta}(k) =
 \frac{(\sin(\theta/2) \sin k,\  \sin(\theta/2) \cos k,\ -\cos(\theta/2) \sin k )}{\sin E_\theta(k)}. 
\end{equation}
A typical band structure $\pm E_{\theta}(k)$
is shown in Fig.\ref{qwpicture}b.
Note that for $\theta_* = 0$ or $2\pi$, the spectrum of $H(\theta_*)$
is gapless and $\vec{n}_{\theta_*}(k_*)$ is ill-defined for $k_*=0,\pi$.

Hamiltonians of the form (\ref{Hk}) can support topological phases if they possess certain symmetries, as indicated in Fig.\ref{TopTable}.
The time-reversal and particle-hole symmetries of this table are defined by the existence of antiunitary operators $\mathcal{T}$ and $\mathcal{P}$ satisfying
\begin{eqnarray}
\label{TRS} \mathcal{T}H\mathcal{T}^{-1} &=& H,\\
\label{PHS} \mathcal{P}H\mathcal{P}^{-1} &=& -H.
\end{eqnarray}
The Hamiltonian $H(\theta)$ given by Eqs.(\ref{Hk}) and (\ref{nk})
possesses PHS (\ref{PHS}) with $\mathcal{P} \equiv K$, where $K$
is the complex conjugation operator.
To see this, note that the evolution operator $U(\theta)$ given by Eqs.(\ref{BasicWalk}-\ref{rotation}) is real, and thus invariant under $K$. 
Along with Eq.(\ref{Heff}), this implies $H^{*}(\theta) = - H(\theta)$, which satisfies Eq. (\ref{PHS}) with $\mathcal{P} \equiv K$. 
In addition, using Eq. (\ref{nk}), it is straightforward to check that $H(\theta)$ possesses a unitary
``chiral'' symmetry 
of the form
\begin{equation}
\label{chiral} \Gamma_\theta^{-1} H(\theta)\Gamma_\theta =- H(\theta),
\end{equation}
with $\Gamma_\theta  = e^{-i\pi \vec{A}_\theta \cdot {\bm
\sigma}/2}$, where $\vec{A}_\theta = (\cos (\theta/2),\, 0,\,
\sin(\theta/2))$ is perpendicular to $\vec{n}_{\theta}(k)$ for all $k$.
The presence of both PHS (\ref{PHS}) and chiral symmetry
(\ref{chiral}) guarantees that $H(\theta)$ is invariant under TRS
(\ref{TRS}) with $\mathcal{T} \equiv \Gamma_\theta\mathcal{P}$,
see Refs.\cite{shinseiclassification1, shinseiclassification2}.

The symmetry classes identified in  Fig.\ref{TopTable} are distinguished by whether the
relevant symmetry operators $\mathcal{T}$ and $\mathcal{P}$ square to $1$ or $-1$.
Because here both $\mathcal{T}^2=1$ and $\mathcal{P}^2=1$, $H(\theta)$ belongs to the class of
Hamiltonians labeled ``SSH.'' 
The corresponding integer-valued topological invariant $Z$ has a simple geometrical
interpretation. Chiral symmetry (\ref{chiral}) constrains
$\mathbf{n}_{\theta}(k) $ to lie on a plane which is
perpendicular to $\mathbf{A}_\theta$, and which contains the origin
(see Fig. \ref{qwpicture}c). Thus, $H(\theta)$
can be characterized by the number of times $\mathbf{n}_{\theta}(k)$ winds
around the origin as $k$ runs from $-\pi$ to $\pi$. Since the winding
number of $\vec{n}_{\theta}(k)$ given by Eq.(\ref{nk}) is 1 for all $\theta
\neq 0,2\pi$, the DTQWs implemented in experiments \cite{coldatomexperiment,ions} simulate the $Z=1$ SSH topological phase.

\begin{figure*}[t]
\includegraphics[width = 17cm]{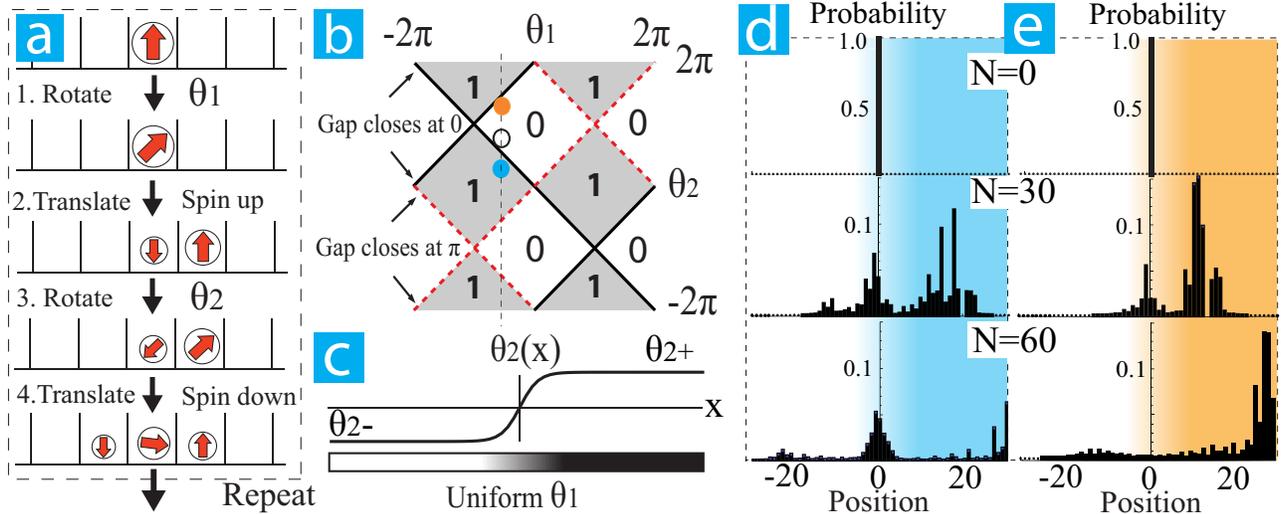}
\caption{(a) One-dimensional ``split-step'' DTQW protocol, see Eq.(\ref{SplitStep}). 
(b) Winding number associated with the split-step DTQW 
as a function of the spin-rotation angles $\theta_{1}$ and $\theta_{2}$.
Topologically distinct gapped phases are separated by phase transition lines where a gap closes at either $E=0$ or $E=\pi$.
(c) Phase boundary in the spatially inhomogeneous split-step DTQW.
In the second rotation stage of Eq.(\ref{SplitStep}), the walker's spin is rotated by an angle $\theta_{2}(x) = \frac12 (\theta_{2-}+ \theta_{2+})+ \frac12(\theta_{2+}- \theta_{2-}) \tanh (x/3)$. 
(d), (e) Dynamics of the spatially inhomogeneous split-step walk, with the walker initialized with spin up at $x = 0$. 
In both panels, we take $\theta_1 = -\pi/2$ and $\theta_{2-} = 3\pi/4$, corresponding to winding number 0 in the region $x \ll 0$ (white dot in panel b).
In (d) we create a phase boundary by taking $\theta_{2+} = \pi/4$, which gives winding number 1 for $x \gg 0$ (see blue dot in panel b).  
After many steps, the probability to find the walker near $x = 0$ remains large, 
indicating the existence of at least one localized state at the phase boundary. For this particular example, numerical diagonalization shows that 
there are three localized states at this boundary. 
In (e), we take $\theta_{2+} = 11\pi/8$ (orange dot in panel b), so that the quantum walk in all regions is characterized by winding number 0.
In this case, the probability to find the walker near $x = 0$ after many steps decays to 0, indicating the absence of a localized state at the boundary. }
\label{rudnerwalk_scheme}
\end{figure*}

The non-trivial topological character of the system can be revealed at a boundary between topologically distinct phases.
To open the possibility
to create such a boundary, we introduce the ``split-step'' DTQW protocol shown in Fig.\ref{rudnerwalk_scheme}a.
Starting from the DTQW defined by Eq.(\ref{BasicWalk}), we split the translations of
the spin-up and spin-down components, and insert an additional spin-rotation $R(\theta_2)$ around the $y$-axis in between:
 \begin{equation}
\label{SplitStep}   U_{ss}(\theta_1, \theta_2) = T_\downarrow R(\theta_2)\, T_\uparrow R(\theta_1),
 \end{equation}
where $T_{\uparrow(\downarrow)}$ shifts the walker to the right (left) by one lattice site if its spin is up (down).

The split-step protocol defines a family of effective Hamiltonians $H_{ss}(\theta_1, \theta_2)$ parametrized by the two spin-rotation angles $\theta_1$ and $\theta_2$.
This family realizes both $Z = 0$ and $Z = 1$ SSH topological phases as displayed in 
Fig.\ref{rudnerwalk_scheme}b, with chiral symmetry (\ref{chiral}) given by $\Gamma_{\theta_1,\theta_2} \equiv \Gamma_{\theta_1}$, $\mathcal{P} = K$, and $\mathcal{T} = \Gamma_{\theta_1}\mathcal{P}$ (see Methods).
Gapped phases with winding numbers $Z = 0$ and $Z = 1$ are separated by phase transition lines where the quasi-energy gap closes at either $E = 0$ or $E= \pm \pi$, as indicated in the figure.

We propose to create a phase boundary in the DTQW by replacing
the second (spatially-uniform) spin rotation $R(\theta_2)$ of Eq.(\ref{SplitStep}) with
a site-dependent spin rotation $R[\theta_2(x)]$, which rotates the walker's spin
through an angle $\theta_2(x)$ about the $y$-axis at each site $x$. 
Specifically, we consider the situation where $\theta_2(x) \rightarrow \theta_{2-}$ for $x \ll 0$ and changes monotonically to $\theta_2(x) \rightarrow \theta_{2+}$ for $x \gg 0$ (see Fig.\ref{rudnerwalk_scheme}c). 
Although this protocol is not translationally-invariant, symmetries (\ref{TRS}-\ref{chiral}) are preserved.
In particular, the system retains the chiral symmetry under $\Gamma_{\theta_1}$ for arbitrary $\theta_2(x)$ as long as $\theta_1$ remains uniform.

When the rotation angles $(\theta_1, \theta_{2+})$ and $(\theta_1, \theta_{2-})$ are chosen to realize
topologically distinct phases with $Z = 0$ and $Z = 1$ in the regions $x \ll 0$ and $x \gg 0$, 
a bound state with energy $0$ or $\pi$ exists near the phase boundary $x=0$ \cite{shinseizeroenergy}. 
The existence of such a bound state is guaranteed by topology, and does not depend on the details of the boundary.
The bound state can be probed by initializing the walker at $x=0$ as demonstrated in Fig.\ref{rudnerwalk_scheme}d. 
Because this initial state has a non-vanishing overlap with the bound state, part of the walker's wavepacket will remain localized near $x=0$.
On the other hand, if the pairs $(\theta_1, \theta_{2+})$,  $(\theta_1, \theta_{2-})$ are chosen to lie 
within the same ``diamond''-shaped region of Fig.\ref{rudnerwalk_scheme}b, then the system can be made spatially uniform through a continuous deformation of the Hamiltonian without closing either gap at $E=0$ or $E= \pi$. 
In this case, there are no topologically protected modes 
localized at the boundary.
For monotonic $\theta_2(x)$, this guarantees that the system does not support {\it any} bound states, and the probability to find the walker at $x = 0$ decays to zero with an increasing number of DTQW steps (see Fig.\ref{rudnerwalk_scheme}e).


With further modifications to the DTQW protocol, each of the topological classes in
1D given in Fig.\ref{TopTable} can be realized (see Supplementary Material).
In addition, as we will now discuss, a straightforward extension of the protocol to a higher-dimensional lattice allows the DTQW to simulate topological phases in two dimensions. 

\section{Topological phases in 2D}
\begin{figure*}[t]
\begin{center}
\includegraphics[width = 17cm]{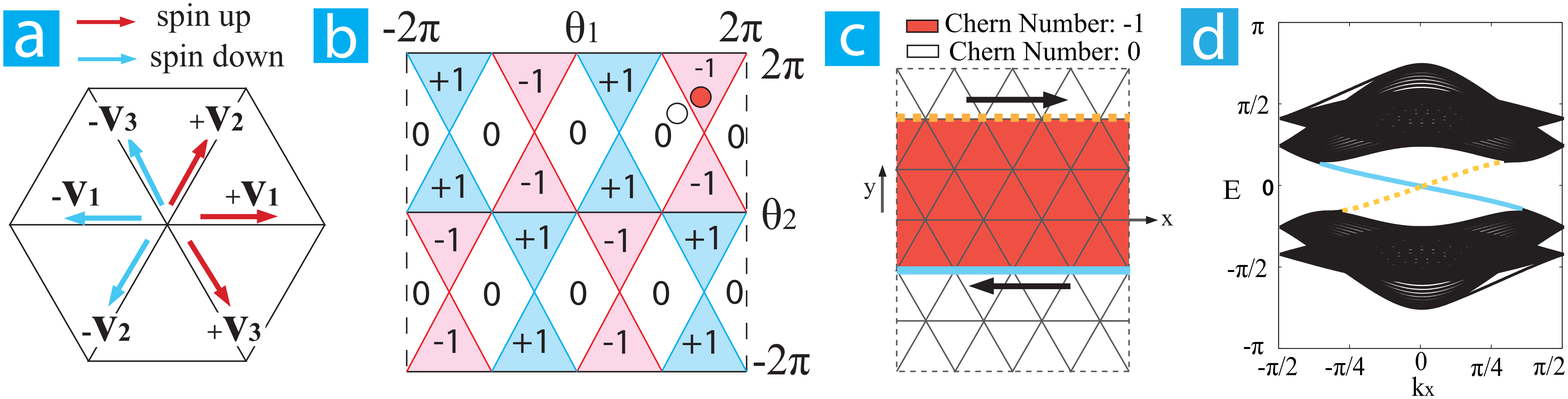}
\caption{(a) Translation vectors for the triangular lattice 2D quantum walk defined in Eq.(\ref{U_triangular}).
(b) Chern number associated with the 2D DTQW as a function of the rotation angles $\theta_1$ and $\theta_2$. 
(c) Geometry of an inhomogeneous 2D quantum walk with periodic boundary conditions. 
In the red region, we take $\theta_{1}=\theta_{2}=3 \pi/2$, corresponding to Chern number $-1$, while in the white region we take $\theta_{1}=\theta_{2}=7\pi/6$, corresponding to Chern number $0$ [see colored dots in panel (b)]. 
Arrows indicate the propagation directions of chiral edge modes localized at the two boundaries.
(d) Quasi-energy spectrum of the inhomogeneous 2D quantum walk depicted in panel (c) for a 100$\times$100 site lattice. The Brillouin zone for momentum $k_x$ parallel to the interface is defined for the doubled unit cell accessed by $U_{\rm 2D}$, Eq.(\ref{U_triangular}). 
Two branches of chiral edge modes connect the upper and lower bands. The dotted (solid) line corresponds to the mode
localized at the upper (lower) boundary in panel (c). }
\label{triangularwalk}
\end{center}
\end{figure*}

To begin, we consider a family of 2D quantum walks in which
the walker possesses two internal states as in the 1D DTQWs above.
Non-trivial topological phases can be realized in a variety of 2D lattice geometries.
Here we consider the case of a triangular lattice, and
discuss equivalent square lattice realizations in the Supplementary Material.
One step of the quantum walk is defined by the unitary operation
\begin{equation}
U_{\rm 2D}(\theta_1,\theta_2)=T_{3}R(\theta_{1})T_{2}R(\theta_{2})T_{1}R(\theta_{1}),
\label{U_triangular}
\end{equation}
where $T_i$ ($i=1,2,3$) translates the walker with spin up
(down) in the $+$($-$)$\mathbf{v}_i$ direction, with
$\{\mathbf{v}_i\}$ defined in Fig.\ref{triangularwalk}a. 
The net result of Eq.(11) is to make the walker hop between sites of a superlattice defined by twice the primitive unit cell.
The effective Hamiltonian for this 2D DTQW takes the form of Eq. (\ref{Hk}) with the integration over $\vec{k} = (k_x, k_y)$ taken over the 2D Brillouin zone (BZ) of the superlattice. 

We now study the topological properties of the 2D DTQWs defined by Eq.(\ref{U_triangular}). 
The corresponding effective Hamiltonians lack time-reversal symmetry, and are thus contained in the symmetry classes in the bottom row of Fig.\ref{TopTable}.
Because $U_{\rm 2D}$ is real, this system possesses PHS with $\mathcal{P} = K$ (see above). 
With a slight modification, this symmetry can be broken and phases in the 
class labeled IQH in Fig.\ref{TopTable} can also be realized (see Supplementary Material).
These phases are analogous to those of the 
Haldane model\cite{haldane}, which exhibits an integer quantum Hall effect in the absence of a net 
magnetic field.

The phases realized by the 2D DTQW, Eq.(\ref{U_triangular}),  
are characterized by an integer-valued topological invariant called the first Chern number.
This quantity is defined in terms
of the unit vector $\mathbf{n}(\mathbf{k})$, see Eq.(\ref{Hk}), as
$C=\frac{1}{4\pi}\int_{BZ} d^2 k [\mathbf{n} \cdot
(\partial_{k_x}\mathbf{n}\times \partial_{k_y}\mathbf{n})]$. 
Geometrically, the Chern number is equal to the number of times $\mathbf{n}(\mathbf{k})$
covers the unit sphere as $\mathbf{k}$ is taken over the 2D
Brillouin zone. We have numerically calculated Chern numbers
for 2D DTQWs throughout the full range
of spin-rotation angles $\theta_1$ and $\theta_2$ (see Methods).
As shown in Fig.\ref{triangularwalk}b, 
phases with $C=0$ and with $C = \pm 1$ can be realized.

Similar to the 1D case, non-trivial topology in 2D DTQWs is manifested
in the presence of protected mid-gap modes bound to the 
interface between two topologically distinct phases. These gapless
modes are analogous to the chiral edge modes of quantum Hall
systems, and are robust against perturbations. 
To confirm the existence of such edge modes, we have used numerical diagonalization
to study a non-uniform 2D DTQW on a 100$\times$100 site triangular lattice with
periodic boundary conditions, see Fig.\ref{triangularwalk}c.
We take the spin-rotation angles $\theta_1$ and $\theta_2$ in Eq.(\ref{U_triangular})
to be site-dependent, with $\theta_1(y) = \theta_2(y) = 3\pi/2$ chosen to realize the
$C = -1$ phase inside the red strip $25 \le y < 75$, and $\theta_1(y) = \theta_2(y) = 7\pi/6$
chosen to realize the trivial $C = 0$ phase outside.
The quasi-energy spectrum is plotted in Fig.\ref{triangularwalk}d as a function of the 
conserved momentum component $k_{x}$ parallel to the interface.
Two counter-propagating chiral edge modes exist inside the bulk gap.
These modes are separately localized at the two boundaries between the $C=0$ and $C=-1$ phases, as indicated in Fig.\ref{triangularwalk}c.

As described above for 1D, these chiral edge modes can be probed by performing the spatially inhomogeneous 2D DTQW described above with the walker initialized at the boundary between two topologically distinct phases. 
Because a general state localized near the phase boundary has a non-zero overlap with the chiral edge mode,
part of the walker's wavepacket will propagate unidirectionally along the boundary.
Such unidirectional propagation is protected by topology, and hence 
is robust even in the presence of an irregularly-shaped boundary.


Finally, we present a time-reversal-invariant 2D DTQW with $\mathcal{T}^2=-1$, which can realize the quantum spin Hall (QSH) phase (see Fig.\ref{TopTable}).
The realization of this phase requires the presence of  
at least four bands, which contain two pairs of time-reversed partners.
Therefore, we now consider a DTQW where the walker possesses four internal states (e.g. a four-level atom, see also experiment \cite{ions}).
We label these four states by a ``spin'' index $\sigma$, which takes the values $\uparrow$ and $\downarrow$, and a ``flavor'' index $\tau$ which takes the values $A$ and $B$.
The time-reversal-invariant unitary step operator $U_{\rm TRI}$ is constructed
in a block-diagonal form,
\begin{equation}
U_{\rm TRI}=\left( \begin{array}{cc} U_A  & 0 \\
0 &\ \ \ U_B
\end{array} \right) \label{TRI}
\end{equation}
where $U_A$ ($U_B$) only acts on the walker if its flavor index is equal to $A$ $(B)$. 
By fixing $U_B=U_A^T$, we ensure that $U_{\rm TRI}$ is invariant under the TRS operation $\mathcal{T}=i\tau_yK$, where $\tau_y$ is a Pauli matrix which acts on the flavor index. 
As an example, if $U_A$ is chosen according to Eq. (\ref{U_triangular}), then
$U_B=R(-\theta_{1})T_{1}^{T}R(-\theta_{2})T_{2}^{T}R(-\theta_{1})T_{3}^{T}$.
Note that $T_{i}^T$ translates the walker in the direction 
$-$($+$)$\mathbf{v}_i$ if its spin is up (down); 
i.e. $T_i^T$ acts opposite to $T_i$.

Time-reversal invariant systems in 2D with $\mathcal{T}^2=-1$ are
characterized by a $Z_2$ topological invariant (middle row of right panel in
Fig.\ref{TopTable}). If $\theta_1$ and $\theta_2$ are chosen such
that $U_A$ is characterized by an odd Chern number, then $U_{\rm
TRI}$ realizes a QSH phase with the $Z_2$ invariant equal to $1$
\cite{topologicalinsulatorprediction2}. Strictly speaking, the
effective Hamiltonian corresponding to $U_{\rm TRI}$ conserves the
flavor index $\tau$ and  
as a result supports topological phases classified by an
integer $Z$, rather than the binary invariant $Z_2$.
However, this additional symmetry can be broken by introducing a coupling
between $A$ and $B$ states which preserves TRS.
In this way, the generic 
$Z_2$ classification can be retrieved (see Supplementary Materials).

\section{Discussion and Summary}
Because the edge modes bound to interfaces between topologically distinct phases in 1D and 2D are topologically protected, their existence is expected to be robust against a broad range of perturbations which may 
arise in real experiments.
In particular, their existence is insensitive to the details of the boundaries, which may be sharp or smooth, straight or curved (in 2D), etc.
In some cases, 
the topological protection arises from certain symmetries (e.g. chiral symmetry in the 1D examples above).
However, even if these symmetries are slightly broken by small errors in the spin-rotation axes and/or angles, 
the edge states are expected to persist due to the absence of nearby states inside the bulk energy gap. 

Throughout this work, we have focused on signatures of topological phases in single-particle dynamics. 
However, some 
dramatic 
manifestations of topological order, e.g. charge fractionalization
and the quantization of the Hall conductivity, appear for specific many-body states 
such as the ``filled-band'' ground states of fermionic systems. 
To observe these phenomena in DTQWs with multiple walkers, analogous many-body states 
can be prepared schematically as follows. 
For special choices of the DTQW parameters, the Bloch eigenstates are simple, i.e. local in space, and uniform in spin. 
By preparing a single filled band comprised of such states, more complicated filled band states can be obtained through a quasi-adiabatic evolution in which the DTQW parameters are changed slightly from step to step.
Even if an energy gap closes along the way, 
the number of excitations created in the process can be controlled by the effective sweep rate. 
In this way, many-body aspects of topological phases may also be studied using DTQWs.

In this paper, we have shown that discrete time quantum walks provide a
unique setting in which 
to realize topological phases in 1D and 2D.
With only slight modifications to the quantum walk protocol which
was realized in recent experiments, the entire ``periodic table''
of topological insulators \cite{shinseiclassification1,
shinseiclassification2} in one and two dimensions can be
explored.
In addition, we have provided a method to detect the presence of topological phases through the appearance of robust edge states at boundaries between topologically distinct phases.

Recently, several promising  system-specific 
methods have been proposed to realize topological phases using cold-atoms
\cite{sorensen, palmer,stanescu,dassarma, Lewenstein, Zoller2, mueller, spielman, dalibard, satija, duan, 
goldman, osterloh}, polar molecules\cite{Zoller}, or photons\cite{fleischhauer}.
Our work advances this emerging field by providing a general framework for studying topological phases in a wide variety of available experimental systems including cold-atoms, trapped ions and photons.
By extending this work to three dimensions, it may be possible to realize
new topological phases, such as the Hopf insulator\cite{wen}, which have not yet been explored in condensed matter systems. 
In addition, multi-particle generalizations of discrete time quantum walks will open new 
avenues in which to explore the quantum many-body dynamics of interacting fermionic or bosonic systems.

\section{Methods}
\subsection{Determination of the phase diagram for 1D split-step DTQW}
The unitary evolution of the 1D split-step DTQW, Eq.(\ref{SplitStep}),
is generated by a Hamiltonian of the form of Eq.(\ref{Hk}) with
$\cos E(k) = \cos(\theta_{2}/2) \cos(\theta_{1}/2) \cos k - \sin(\theta_{1}/2)\sin(\theta_{2}/2) $,
and
\begin{eqnarray} \label{splitstepn}
n_{x}(k) &=&  \frac{\cos(\theta_{2}/2) \sin(\theta_{1}/2) \sin k}{\sin E(k)} \nonumber \\
n_{y}(k) &=&  \frac{\sin(\theta_{2}/2)\cos(\theta_{1}/2)+ \cos(\theta_{2}/2)\sin(\theta_{1}/2) \cos k}{\sin E(k)} \nonumber \\
n_{z}(k) &=&  \frac{-\cos(\theta_{2}/2) \cos(\theta_{1}/2) \sin k }{\sin E(k)}  \nonumber
\end{eqnarray}
It is straightforward to check that $\vec{A}(\theta_{1})= (\cos(\theta_{1}/2),\, 0,\, \sin(\theta_{1}/2))$
is perpendicular to $\vec{n}(k)$ for all $k$.
Therefore, the system possesses chiral symmetry (\ref{chiral})
with $\Gamma(\theta_{1})= e^{-i\pi \vec{A}(\theta_{1}) \cdot {\bm \sigma}/2}$.
As a result, the split-step DTQW can be characterized by the winding number of
$\vec{n}(k)$ around the origin, denoted by $Z$. Using the explicit expression for $\vec{n}(k)$ in Eq. (\ref{splitstepn}), we find $Z = 1$ if
$|\tan(\theta_{2}/2)/\tan(\theta_{1}/2)| <1$, and $Z = 0$ if $|\tan(\theta_{2}/2)/\tan(\theta_{1}/2)| >1$.
The spectrum is gapless along the lines $|\tan(\theta_{2}/2)/\tan(\theta_{1}/2)| = 1$. 
Thus we obtain the phase diagram displayed in Fig. \ref{rudnerwalk_scheme} (b).

\subsection{Localized states at a phase boundary of inhomogeneous split-step 1D DTQW}
In addition to the dynamical simulations presented in the main text, we have confirmed the existence of topologically protected edge states  with energy $E=0$ or $E=\pi$ 
in the 1D split-step DTQW through an analytical calculation for an infinite system with a {\it sharp} boundary,
using $\theta_{2}(x) = \theta_{2-}$ for $x<0$ and $\theta_{2}(x) = \theta_{2+}$ for $x \ge 0$.
Furthermore, we have used numerical diagonalization to study the spectrum of a finite (periodic) system on a ring which hosts two phase boundaries.
In all cases, we find that if the phases on the two sides of a boundary are topologically distinct, i.e. characterized by different winding numbers $Z$, then a single localized state with energy $E=0$ or $E= \pi$ exists at the boundary.

For {\it smooth} boundaries as described in the main text,  other localized states that are not 
protected by topology could appear. 
These bound states always appear in pairs with energies $E$ and $-E$ due to 
chiral symmetry. 
Therefore, when the phases on the two sides of a boundary are topologically distinct, 
an odd number of bound states appears at the phase boundary\cite{shinseizeroenergy}.

\subsection{Phase diagram of the 2D DTQW}
Here we briefly describe a general procedure for determining the phase diagrams of 2D DTQWs. 
Because the value of a quantized topological invariant can only change across a phase boundary where a gap closes, we first identify the lines in parameter space along which a gap vanishes in the quasi-energy spectrum.
Once these phase boundaries are determined, the topological phases between boundaries can be identified by computing the topological invariant at any single point within each region. 
For the 2D DTQW with evolution operator given by Eq.(\ref{U_triangular}), we have 
obtained phase boundaries analytically from the spectrum 
\begin{eqnarray*}
\cos E(\vec{k})\!\!\!\!\! &&= \left\{\cos(\theta_{2}/2) \cos\theta_{1} \cos( \mathbf{k} \cdot (\mathbf{v}_{1}+ \mathbf{v}_{2})) \right. \\
 && \left.- \sin(\theta_{2}/2) \sin\theta_{1} \cos( \mathbf{k} \cdot (\mathbf{v}_{1}- \mathbf{v}_{2}) \right\}
\cos(\mathbf{v}_{3} \cdot \mathbf{k}) \\
& & - \cos(\theta_{2}/2) \sin( \mathbf{k} \cdot (\mathbf{v}_{1}+ \mathbf{v}_{2}))
\sin(\mathbf{v}_{3} \cdot \mathbf{k}),
\end{eqnarray*}
which gives the lines shown in Fig. \ref{triangularwalk} (b). 
We then numerically evaluated the Chern number within each region using $C=\frac{1}{4\pi}\int_{BZ} d^2 k [\mathbf{n} \cdot
(\partial_{k_x}\mathbf{n}\times \partial_{k_y}\mathbf{n})]$ with the appropriate expression for $\vec{n}(\vec{k})$.

\section{acknowledgements}
We are grateful to Y. Shikano for introducing us to DTQWs.  
We thank M. D. Lukin and M. Levin for useful discussions. This work is
supported by NSF grant DMR 0705472,  CUA, DARPA OLE, AFOSR MURI. 
E.B. was also supported by the NSF under grants DMR-0757145, 
and MSR was supported by NSF Grants DMR 090647 and PHY 0646094.

\newpage \clearpage

\part{Online Supplementary Material}

\section{Explicit DTQW protocols for all topological classes}
In this section, we provide explicit DTQW protocols which can be used to realize topological phases in each of the symmetry classes listed in Fig.1 of the main text.
These protocols are summarized in Supplementary Fig.\ref{table}.
Each DTQW presented in Fig.\ref{table} can realize both trivial and non-trivial phases within a given symmetry class.
The specific phase which is realized is determined by the spin rotation angles which parametrize the quantum walk;
the system can be driven through a topological phase transition by tuning these spin rotation angles.
In the following, we denote the presence of time-reversal symmetry (TRS) with $\mathcal{T}^2 = \pm 1$
by TRS$=\pm1$, and the absence of time-reversal symmetry by TRS$=0$.
Similarly, we denote the presence of particle-hole symmetry (PHS) with $\mathcal{P}^2 = \pm 1$ by PHS$=\pm 1$, and its absence by PHS$=0$.
We denote the presence of chiral symmetry under the unitary operator $\Gamma$ by CH=1, and its absence by CH=0.
Note that because the chiral symmetry operator $\Gamma$ is {\it unitary}, rather than {\it antiunitary}, the phase of its square does not carry any additional information.  
In particular the transformation $\Gamma \rightarrow e^{i\theta}\Gamma$ results in $\Gamma^2 \rightarrow e^{2i\theta}\Gamma^2$. 
For all of the DTQWs considered below,
the presence of any two of the symmetries \{TRS, PHS, CH\} automatically
ensures the presence of the third.
For example, if a system possesses PHS and
CH under the operators $\mathcal{P}$ and $\Gamma$, then it also
possess TRS under the operator $\mathcal{T} = \mathcal{P}\Gamma$.

\begin{figure*}[t]
\begin{center}
\includegraphics[width = 17cm]{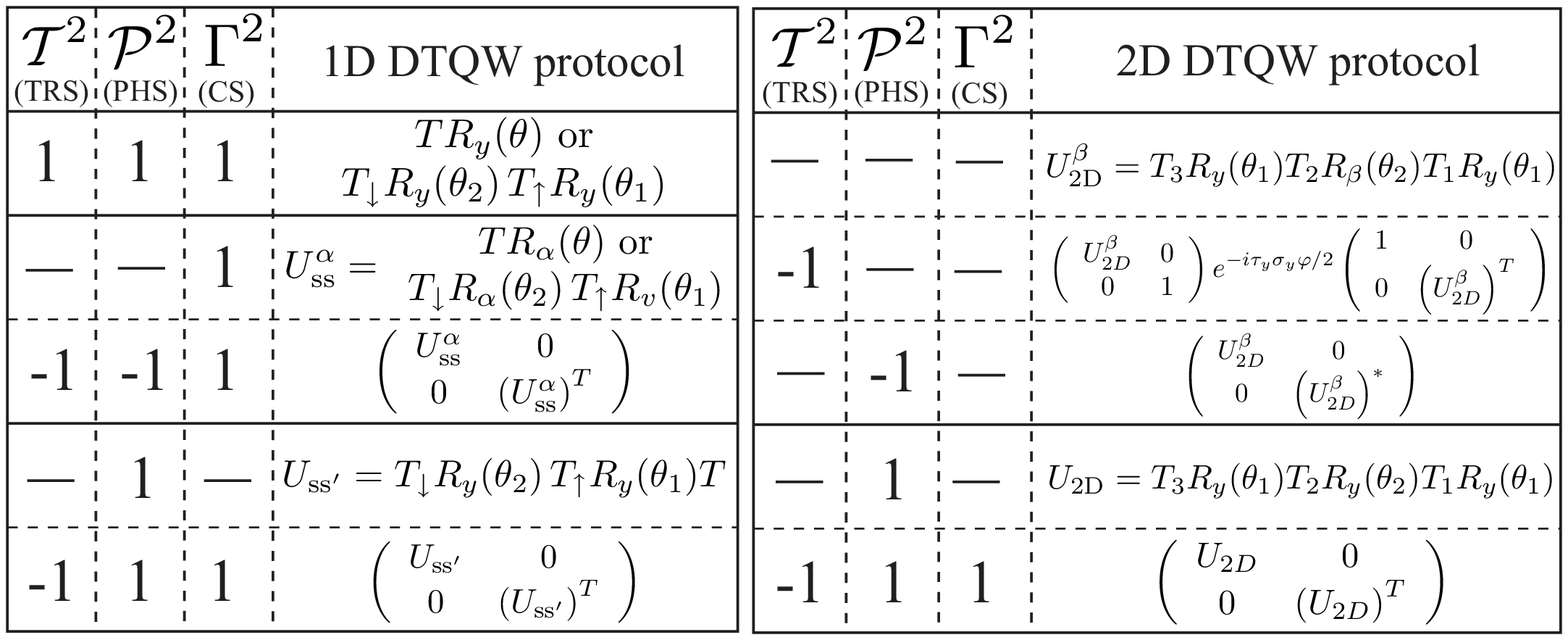}
 \caption{DTQW protocols for each symmetry class of topological phases in 1D and 2D.
By tuning the rotation angles, all of these examples can realize
both trivial and non-trivial topological phases within each class.
Here $T$ translates the walker to the right (left) if its spin is up (down), while $T_{\uparrow} (T_{\downarrow})$ translates only the spin up (down) component to the right (left).
In 2D, the translation $T_{i}$ shifts the walker in the $\vec{v}_{i}$ ($-\vec{v}_{i}$)
direction if its spin is up (down), see Fig.4a of the main text.
A spin rotation operator $R_u(\theta)$ rotates the walker's spin through an angle $\theta$ about the axis $u \in \{y, \alpha, \beta\}$, where $\mathbf{\alpha} = \frac{1}{\sqrt{2}} (0,1,1)$, and $\vec{\beta} = (\sin(\pi/8),\cos(\pi/8),0)$.
In most cases, quantum walks with 
TRS=$-1$ and PHS=-1 are obtained by the doubling procedure starting from a quantum walk with evolution operator $U_A$ which has TRS=0 and PHS=0. 
Such cases are separated by dotted lines.
See main text for descriptions of the relevant symmetry operators.}
\label{table}
\end{center}
\end{figure*}
\vspace{-0.15 in}

\subsection{Doubling Procedure}
Quantum walks with TRS$=-1$ can be readily constructed from DTQWs
with TRS=0 through the ``doubling procedure'' used to construct
$U_{\mathrm{TRI}}$[Eq.(12)] in the main text. First the walker is
endowed with an additional two-fold ``flavor'' index $\tau$ which
can take either the value $A$ or $B$. We then choose an evolution
operator which is diagonal in the flavor index and which satisfies
$U_B = U_A^T$, where $U_{A(B)}$ is the evolution operator which
acts on the walker with flavor $A$($B$).
With this possibility in mind, below we focus on examples with TRS=0.

\vspace{-0.1 in}

\subsection{One dimensional topological phases}

\vspace{-0.1 in}
\subsubsection{{\bf Symmetry Classes:}\\TRS$=0$, PHS$=0$, CH$=1$ ($Z$) \\ TRS$=-1$, PHS$=-1$, CH$=1$ ($Z$)}
The split-step DTQW described by Eq.(10) of the main text realizes the symmetry class with TRS=$1$, PHS=$1$, CH=$1$.
By an appropriate change of the direction of the spin-rotation axes, 
TRS and PHS can be broken while CH is retained.
Thus in order to realize the related symmetry class TRS=$0$, PHS=$0$, CH=$1$, we will break the PHS of the split-step DTQW.

In the main text we showed that any DTQW whose unitary evolution operator is {\it real} possesses PHS with $\mathcal{P} = K$, where $K$ is the complex conjugation operator.
The existence of PHS is in fact more general: if the two spin-rotations in a
split-step DTQW are performed around the same axis, and if that axis lies in the
$xy$-plane, then the DTQW will have PHS=1.
To see this, suppose that both rotations are performed around the axis
$(\sin\phi,\cos\phi,0)$. It is then straightforward to check that the resulting effective Hamiltonian possesses PHS under the operator $\mathcal{P}
= e^{-i \sigma_{z} \phi/2}K e^{i \sigma_{z} \phi/2}$.

On the other hand, PHS is absent if we choose a rotation axis that contains
a non-zero $z$ component. An example of a DTQW with PHS=0 is
provided by the evolution operator
 \begin{equation}
  U^{\alpha}_{\mathrm{ss}}(\theta_1, \theta_2) =
  T_\downarrow R_{\alpha}(\theta_2)\, T_\uparrow R_{\alpha}(\theta_1),\label{Uv}
 \end{equation}
where $R_{\alpha}(\theta)$ is a spin-rotation around the axis
$\mathbf{\alpha} = \frac{1}{\sqrt{2}} (0,1,1)$ through the angle $\theta$.
Although PHS is absent, this system possesses chiral symmetry under the symmetry operator
$\Gamma_{\alpha}(\theta_{1}) = i e^{-i\pi \vec{A}_{\alpha}(\theta_{1}) \cdot{\bm \sigma}/2}$, where
$\vec{A}_{\alpha}(\theta_{1})= (\cos(\theta_{1}/2), \frac{1}{\sqrt{2}}\sin(\theta_{1}/2), \frac{1}{\sqrt{2}}\sin(\theta_{1}/2) )$.
The absence of TRS can be checked in the following way. If the energy eigenvalues of the two states with momentum $k$ are given by $\pm |E(k)|$, 
then TRS$=\pm1$ requires $|E(k)| = |E(-k)|$. 
We have explicitly checked that this relation is not satisfied for the DTQW defined by Eq.(\ref{Uv}), and thus conclude that TRS is absent.


The split-step DTQW above, Eq.(\ref{Uv}), can realize distinct topological phases by tuning the spin-rotation
angle $\theta_{2}$. For example, the trivial phase with winding number $Z = 0$ is realized with $\theta_{1}= \pi/2$ and $\theta_{2}= 3\pi/4$ and the phase with winding number $Z = 1$ is realized with  $\theta_{1}= \pi/2$ and $\theta_{2}= \pi/4$.

The recent experimental implementation of a DTQW with photons\cite{photons} employed the rotation operator given by the 
Hadamard gate $R_{\mathcal{H}} = i e^{-i \pi \vec{n} \cdot {\bm \sigma}/2}$, with $\vec{n} = 1/\sqrt{2}(1,0,1)$. Since the
rotation axis contains a non-zero $z$ component, we conclude that this ``Hadamard walk'' belongs to the symmetry class TRS=$0$, PHS=$0$, CH=$1$.

Using the doubling procedure described above, a time-reversal
symmetric DTQW with TRS=$-1$, PHS=$-1$, and CH=$1$ can be
constructed based on the DTQW defined in Eq. (\ref{Uv}). The
corresponding evolution for one step of the DTQW is given by ${\rm
diag}\big[ U^{\alpha}_{\mathrm{ss}}(\theta_1, \theta_2),\, \left(
U^{\alpha}_{\mathrm{ss}}(\theta_1, \theta_2)\right)^{T}\big]$. It
is straightforward to check that this quantum walk possesses
chiral symmetry under the operator $\Gamma
=\mathrm{diag}\big[\Gamma_{\alpha}(\theta_{1}),
\Gamma_{\alpha}^{*}(\theta_{1}) \big]$. By construction, this DTQW
possesses TRS$=-1$ with $\mathcal{T} = i \tau_{y} K$. Using these
two symmetries, we construct a PHS operator $\mathcal{P} = \Gamma
\mathcal{T}$ with $\mathcal{P}^2 = -1$.


\subsubsection{{\bf Symmetry Classes:}\\TRS=$0$, PHS=$1$, CH=$0$ ($Z_{2}$) \\TRS=$-1$, PHS=$1$, CH=$1$($Z_{2}$)}
The construction of a DTQW with TRS=$0$, PHS=$1$, CH=$0$ starts
from the split-step DTQW with TRS=$1$, PHS=$1$, CH=$1$ [see main text, Eq.
(10)]. The chiral symmetry 
can be broken by adding extra operations to the split-step DTQW.
On the other hand, in the previous section we showed that PHS can
be retained quite generally as long as the two rotation axes are the same
and taken to lie on $xy$-plane.

In order to construct a DTQW with CH=$0$, we begin with Eq.(10) 
and add an additional spin-dependent translation $T$ 
which translates the walker to the right(left) by one
lattice site if its spin is up(down) [see Eq.(2)].
Explicitly, the evolution operator for one step of a representative DTQW from this symmetry class is given by
 \begin{equation}
  U_{\mathrm{ss'}}(\theta_1, \theta_2) = T_\downarrow R_y(\theta_2)\, T_\uparrow R_y(\theta_1)
  T,\label{Uch0}
 \end{equation}
 where $R_y(\theta)$ is a spin-rotation around the $y$ axis through an angle $\theta$ [Eq.(3)].
 Since $U_{\mathrm{ss'}}$ is real, this DTQW retains PHS=1 with $\mathcal{P} = K$.
The absence of chiral symmetry for this walk can be verified by observing
that the quantization axis $\vec{n}(k)$ does not lie on a plane which includes the origin.
Therefore, no single operator $\Gamma$ can be found which satisfies  $\Gamma H(k)=-H(k)\Gamma$ for all $k$.

One dimensional systems with particle-hole symmetry exhibit two distinct topological phases\cite{Qi1}.
These two phases are indexed by the Berry phase, which can only take the quantized values $0$ and $\pi$
due to the presence of PHS. Explicitly, the invariant is given by
\begin{eqnarray}
B = \int \frac{dk}{2\pi} (-i) \bra{\psi_{lb}(k)}  \partial_{k} \ket{\psi_{lb}(k)}.
\end{eqnarray}
Here, $\ket{\psi_{lb}(k)}$ is the eigenstate in ``lower band'' with momentum $k$.
 The DTQW described above can realize both topological phases, with the trivial phase ($B=0$) realized for
 $\theta_{1} = \pi/2, \theta_{2} = \pi/6$, and
 the non-trivial phase with $B=1/2$ realized for $\theta_{1} = \pi/2, \theta_{2} = 2\pi/3$.

Using the doubling procedure, we can construct a time-reversal invariant DTQW with TRS=$-1$, PHS=1, CH=1 based on Eq. (\ref{Uch0}). 

\newpage
\subsection{Two dimensional topological phases}
\subsubsection{{\bf Symmetry Classes:}\\TRS=$0$, PHS=$1$, CH=$0$ ($Z$)\\ TRS=$-1$, PHS=$1$, CH=$1$($Z_{2}$)}
The triangular lattice 2D DTQW defined by Eq.(11) of the main text involves only spin
rotations around the $y$-axis.
Consequently, the evolution operator $U_{\rm 2D}$ is real and possesses PHS=$1$ with $\mathcal{P} = K$.
Therefore, the time-reversal invariant DTQW $U_{\mathrm{TRI}}$ constructed from $U_{2D}$, Eq.(12), is contained in the symmetry class
TRS=$-1$, PHS=$1$, CH=$1$.

As noted in the main text, $U_{\mathrm{TRI}}$
is diagonal in the flavor index $\tau = A,B$ and thus possess an
extra symmetry related to the conservation of $\tau_{z}$.
Here we describe a more general 2D DTQW with TRS=$-1$ which does
not possess this additional symmetry.
The operator for one step of this modified time-reversal invariant DTQW is given by
\begin{equation}
U_{\rm TRI'}=\left( \begin{array}{cc} U_A  & 0 \\ 0 &  1 \end{array} \right)
e^{-i\tau_{y}\sigma_{y} \phi/2}
\left( \begin{array}{cc} 1  & 0 \\ 0 &U_{B} \end{array} \right)
\label{TRIp}
\end{equation}
where $U_{A}(U_{B})$ acts on the walker if its flavor index is
$A(B)$. The rotation 
$e^{-i\tau_{y}\sigma_{y} \phi/2}$ explicitly introduces mixing
between the $A$ and $B$ flavors, and thus breaks the conservation of
$\tau_z$.

This DTQW is characterized by TRS=$-1$ with the symmetry operator
$\mathcal{T} = i \tau_{y} K$ if $U_{B}$ is chosen according to
$U_{B} =U_{A}^{T} $. If the Chern number associated with $U_{A}$
is odd, then $U_{\rm TRI'}$ with $\phi=0$ realizes a non-trivial
QSH topological phase. Because this phase is protected by TRS, the
presence of a small $\phi>0$ can not take the system out of this
phase.

\subsubsection{{\bf Symmetry Classes:}\\TRS=$0$, PHS=$0$, CS=$0$ ($Z$)\\ TRS=$-1$, PHS=$0$, CH=$0$ ($Z_{2}$)}
The existence of topological phases characterized by a non-zero
Chern number does not rely on the presence of PHS. Therefore, the
topological phase with Chern number $1$ in the TRS=$0$, PHS=$1$,
CS=$0$ symmetry class can be directly transformed to the corresponding phase in the TRS=$0$, PHS=$0$, CS=$0$ symmetry class
by a perturbation which breaks PHS.
Such a perturbation can be achieved by changing the rotation axis for the second
rotation stage in Eq.(11). The resulting DTQW single-step evolution operator is given by
\begin{equation}
U^{\beta}_{\mathrm{2D}}(\theta_1,\theta_2)=T_{3}R(\theta_{1})T_{2}R_{\beta}(\theta_{2})T_{1}R(\theta_{1}),
\label{U_2D_p}
\end{equation}
where $R$ is a spin rotation around the $y$ axis, and
$R_{\beta}(\theta)$ is a spin rotation around the axis $\mathbf{\beta} = (\sin\phi,\cos\phi,0)$ with $\phi = \pi/16$.
The operators $\{T_{i}\}$ correspond to spin-dependent translations along the directions $\{\vec{v}_i\}$, as
defined in Fig.4a.
The absence of PHS is confirmed by examining the relationship between energy eigenvalues 
$|E(\mathbf{k})|$ and $|E(-\mathbf{k})|$. The presence of PHS implies $|E(\mathbf{k})| = |E(-\mathbf{k})|$. 
This condition is violated for DTQW (\ref{U_2D_p}). Therefore this system does not possess PHS.

DTQW (\ref{U_2D_p}) realizes both topologically trivial and non-trivial
phases with zero and nonzero Chern numbers. 
For example, the choice $\theta_{1}= \theta_{2}= 3\pi/2$ generates
the phase with Chern number $-1$, while $\theta_{1}= \theta_{2}=
7\pi/6$ corresponds to the phase with Chern number $0$. Since PHS
is absent, this DTQW belongs to the class with TRS=0, PHS=0, CS=0.
The related time-reversal invariant DTQW constructed by applying
the doubling procedure to this walk has TRS$=-1$, PHS=0, CH=0.

\subsubsection{{\bf Symmetry Class:}\\TRS=0, PHS=$-1$, CS=0 ($Z$)}
Quantum walks with PHS=$-1$ can be constructed through a doubling procedure similar to that used to construct DTQWs with TRS=$-1$.
Consider the block-diagonal evolution operator
\begin{equation}
U_{\rm PHI}=\left( \begin{array}{cc} U_A  & 0 \\ 0 &  U_{B}
\end{array} \right), \label{PHI}
\end{equation}
where $U_{A}(U_{B})$ acts on the walker if its flavor index is
$A(B)$. If we choose $U_{B}=U^{*}_{A}$, then the resulting DTQWs 
possess PHS=$-1$ with $\mathcal{P} = i \tau_{y} K$.
By choosing $U_{A}$ according to Eq.(\ref{U_2D_p}) with parameters to give a 
Chern number of $1$, 
Eq. (\ref{PHI}) produces a DTQW which realizes a non-trivial topological
phase in the symmetry class TRS=0, PHS=$-1$, CH=0.

\section{Realization of 2D topological phases on a square lattice}
In the main text, we have provided examples of 2D DTQWs which realize topological
phases on a triangular lattice.
However, 
these DTQWs 
can also be implemented on a square lattice, as we explain below.
A square lattice may be easier to realize in some experimental
implementations, such as cold atoms in optical lattices. 

Generally speaking, the phase diagram of a DTQW is determined by
the amplitude for the walker to hop from one site to another after
one complete step of the evolution.
Thus, as long as the hopping amplitudes between all pairs of sites are preserved, 
geometrical deformations of the lattice do not change the phase
diagram. 
Therefore, the phase diagram of a DTQW is insensitive to geometric deformations of its host lattice.

\begin{figure}[t]
\begin{center}
\includegraphics[width = 4cm]{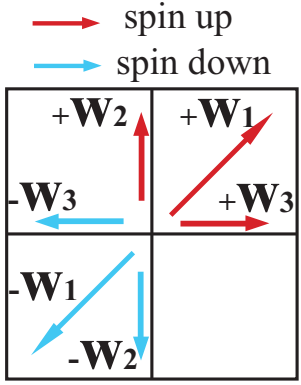}
\caption{Translation vectors for 2D DTQW on a square lattice that realizes the phase diagram
of Fig.4 in the main text. Crucially, $\vec{v}_{3}$ satisfies the relation
$\vec{v}_{3} =\vec{v}_{1} - \vec{v}_{2}$ just as in the triangular lattice realization.}
\label{squarelattice}
\end{center}
\end{figure}

In particular, 2D DTQWs with non-zero Chern number can be realized on
a square lattice by replacing the translations along the vectors $\{\vec{v}_i\}$ on the triangular lattice in Eq.(11) with the vectors $\{\vec{w}_{i}\}$ shown in Fig.\ref{squarelattice}.
Here, $\vec{w}_1=(1,1)$, $\vec{w}_2=(0,1)$ and $\vec{w}_3=(1,0)$.
This protocol is obtained simply by ``shearing'' the lattice used in Eq.(11) and Fig.4a.
Note that the diagonal translation along $\vec{w}_1$ can be implemented by a compound translation along $(1,0)$ followed by a translation along $(0,1)$.

\end{document}